\documentclass[conference]{IEEEtran}
\IEEEoverridecommandlockouts
\usepackage{cite}
\usepackage{amsmath,amssymb,amsfonts}
\usepackage{graphicx}
\usepackage[caption=false,font=footnotesize]{subfig}
\usepackage{booktabs}
\usepackage{url}
\usepackage{textcomp}
\usepackage[table]{xcolor}
\usepackage{tikz}
\usetikzlibrary{arrows.meta,decorations.pathmorphing,calc}
\graphicspath{{fig/}}

\newcommand{\best}[1]{\textbf{#1}}
\newtheorem{theorem}{Theorem}
\newtheorem{remark}{Remark}
\newtheorem{proposition}{Proposition}
\DeclareRobustCommand{\circled}[1]{\tikz[baseline=(c.base)]\node[circle,fill=black!75,text=white,font=\tiny\bfseries,inner sep=0.6pt,minimum size=2.6mm](c){#1};}
    
\begin{document}

\title{Modal--Augmented Chebyshev Physics--Informed Kolmogorov--Arnold Networks for Power Systems\\
\thanks{The work was partially supported by the Estonian Centre of Excellence in Energy Efficiency, ENER (grant TK230) funded by the Estonian Ministry of Education and Research. Code and data: \protect\url{https://github.com/komeilnosrati/mac-PIKAN}, komeil.nosrati@taltech.ee}
}

\author{
\IEEEauthorblockN{Komeil Nosrati, Majid Ghorbani, Aleksei Tepljakov, Eduard Petlenkov}
\IEEEauthorblockA{
\textit{Department of Computer Systems}\\
\textit{Tallinn University of Technology}\\
Tallinn, Estonia\\}
\and
\IEEEauthorblockN{Juri Belikov}
\IEEEauthorblockA{
\textit{Department of Software Science}\\
\textit{Tallinn University of Technology}\\
Tallinn, Estonia}
}

\maketitle

\begin{abstract}
Physics-informed neural networks (PINNs) offer a promising alternative to numerical integration for power system dynamic simulation; however, their conventional architectures can struggle to capture electromechanical oscillations. Although physics--informed Kolmogorov--Arnold networks (PIKANs) offer an alternative, their B-spline implementations incur substantial training costs. To address these limitations, this paper investigates Chebyshev PIKANs (c--PIKANs) and proposes a modal--augmented c--PIKAN (mac--PIKAN) based on a trial solution that incorporates the closed-form modal response of the linearized swing equation. First, the modal response is used to capture the dominant oscillatory dynamics. Then, a c--PIKAN learns a nonlinear correction weighted by a quadratic time-dependent factor, ensuring that the initial rotor angle and speed are satisfied exactly. Evaluations on the single-machine infinite-bus benchmark demonstrate the computational efficiency of c--PIKAN and show that mac-PIKAN reduces PINN rotor--angle and frequency errors from 1.59\% to 0.29\% and from 5.26\% to 0.99\%, respectively, while maintaining higher accuracy across training-data sizes and at lower damping levels.
\end{abstract}


\section{Introduction}
Physics--informed neural networks (PINNs) have attracted growing interest as an alternative to conventional numerical integration for power system dynamic simulation, including rotor--angle prediction~\cite{raissi2019pinn}, inertia and damping estimation~\cite{misyris2020pinn,stiasny2021sysid}, and accelerated simulation~\cite{stiasny2023tds,venturanadal2025}. However, standard PINNs use multilayer perceptrons (MLPs), whose spectral bias towards smooth functions limits their ability to capture electromechanical oscillations accurately~\cite{xu2026review,krishnapriyan2021failure}. Kolmogorov--Arnold networks (KANs)~\cite{liu2024kan}, which represent nonlinearities through learnable edge functions, offer a promising alternative. Building on this concept, Shuai and Li introduced physics--informed KANs (PIKANs) with B-spline basis functions for power system dynamics, reporting improved accuracy with fewer parameters~\cite{shuai2025pikan}. However, their computational cost motivates more efficient architectures~\cite{toscano2025pikan}, fair comparisons with the PINN under common settings, and physical constraints for exact initial-condition enforcement.

\begin{figure*}[t]
\centering
\resizebox{0.99\textwidth}{!}{\includegraphics{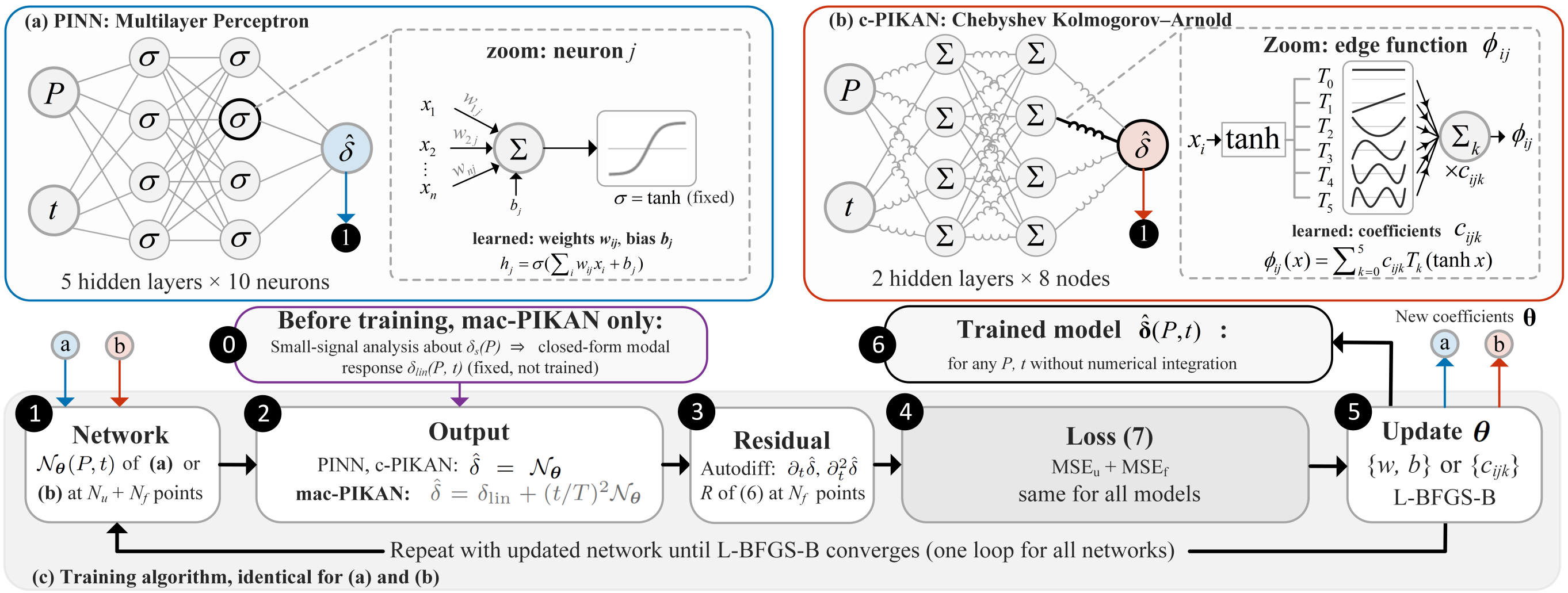}}
\caption{(a) PINN: scalar weights and a fixed $\tanh$ in every neuron; (b) c--PIKAN: one learnable Chebyshev function per edge, nodes only sum; (c) Training, identical for both: \circled{1} network, \circled{2} output, \circled{3} residual \eqref{eq:res}, \circled{4} loss \eqref{eq:loss}, \circled{5} parameter update, \circled{6} trained model; mac--PIKAN adds \circled{0} closed-form small-signal response \eqref{eq:zlin} to the output \eqref{eq:ps}, so both initial conditions hold exactly (with the MLP it is ma-PINN (see Remark~\ref{rem:interpretation})).}
\label{fig1}
\end{figure*}

Chebyshev-based PIKANs (c--PIKANs) offer a computationally efficient alternative to B--spline--based architectures through polynomial representations, with competitive benchmark performance~\cite{shukla2024fair,toscano2025pikan}. However, their effectiveness in power system dynamic simulation remains largely unexplored~\cite{shuai2025pikan}. Further improvements can exploit the physical structure of power system dynamics~\cite{xu2026review}. In particular, small--signal analysis provides a closed-form modal response describing dominant electromechanical oscillations around an operating point. Embedding this response into the trial solution enforces initial conditions exactly and reduces learning to the nonlinear deviation from linearized dynamics, without additional parameters~\cite{lagaris1998}. Accordingly, this work investigates c--PIKANs for power system dynamic simulation and proposes a modal-augmented c--PIKAN (mac--PIKAN), combining the linearized swing-equation modal response with a trainable correction that vanishes initially, as illustrated in Fig.~\ref{fig1}. The main contributions are:
\begin{itemize}
    \item A c--PIKAN formulation of the swing equations that replaces B--spline basis functions with Chebyshev polynomials to reduce the training cost of PIKANs.
    \item A mac--PIKAN framework with theoretical guarantees that enforces initial conditions exactly and learns only a second-order correction to the small-signal response.
    \item A comparative evaluation of baseline methods and the proposed modal-augmented variant, focusing on initial--speed constraints and robustness to damping variations.
\end{itemize}
Under the benchmark setting of~\cite{misyris2020pinn}, mac--PIKAN outperforms PINNs across all random seeds and B--spline PIKANs (bs--PIKANs) on average, with largest gains near boundary trajectories. Unlike c--PIKAN, which predicts incorrect initial speeds, mac--PIKAN keeps the initial-speed condition. Reduced damping degrades both methods, yet mac--PIKAN retains its advantage, showing robustness to damping variations.

\section{Problem Formulation}
\begin{figure}[!t]
\centering
\includegraphics[width=\columnwidth]{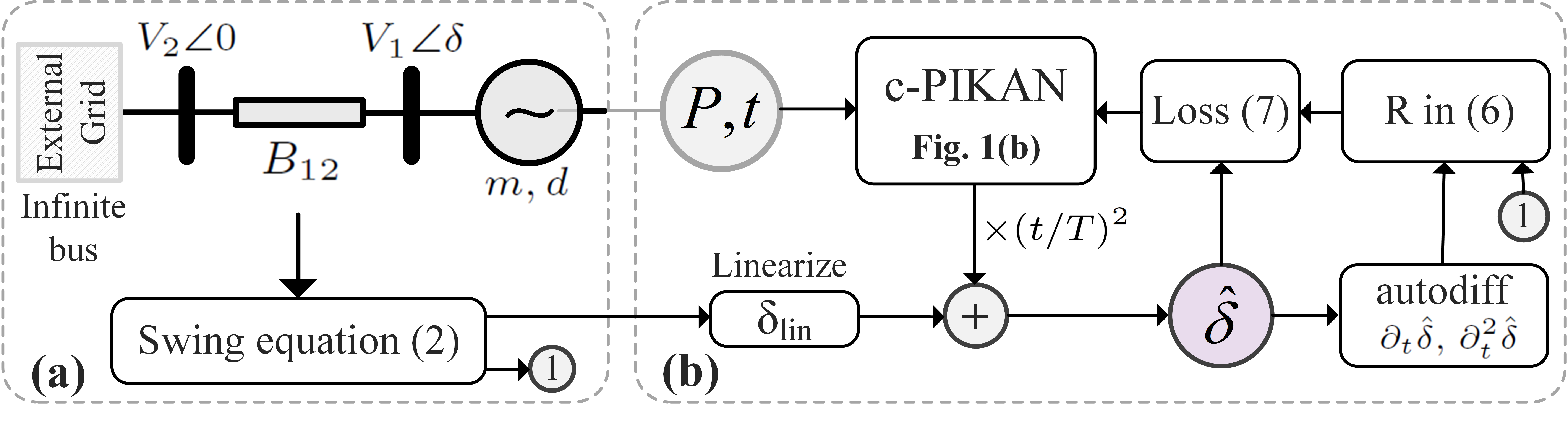}
\caption{(a) SMIB system; (b) general structure of a mac-PIKAN.}
\label{fig:smib}
\end{figure}

Consider an $n$--machine power system with constant internal voltages $E_i$ and constant--admittance loads, governed by the swing equations~\cite{kundur1994}
\begin{equation}
M_i\ddot{\delta}_i+D_i\dot{\delta}_i=P_{m,i}-\sum_{j=1}^{n}E_iE_j\left(G_{ij}\cos\delta_{ij}+B_{ij}\sin\delta_{ij}\right),
\label{eq:mm}
\end{equation}
for $i=1,\ldots,n$, where $\delta_i$ is the rotor angle, $M_i$ and $D_i$ are the inertia and damping coefficients, and $P_{m,i}$ is the mechanical input power. Moreover, $\delta_{ij}=\delta_i-\delta_j$, and $G_{ij}+jB_{ij}$ is the $(i,j)$ entry of the reduced admittance matrix. The summation defines the electrical output power $P_{e,i}(\boldsymbol{\delta})$, with $\boldsymbol{\delta}=[\delta_1,\ldots,\delta_n]^\top$. The single--machine infinite--bus (SMIB) benchmark of~\cite{misyris2020pinn} is the special case, given as
\begin{equation}
m\,\ddot{\delta}+d\,\dot{\delta}+B\sin\delta=P,\qquad \delta(0)=\delta_0,\quad \dot{\delta}(0)=\omega_0,
\label{eq:smib}
\end{equation}
where $m$, $d$, and $P$ are the inertia, damping, and mechanical power, and $B=B_{12}V_1V_2$ is the synchronizing--power coefficient, with $V_1$ and $V_2$ denoting the generator and infinite--bus voltage magnitudes, and $B_{12}$ the line susceptance (see Fig.~\ref{fig:smib}(a)). With initial rotor angle and speed $\delta_0$ and $\omega_0$, the objective is to learn a neural surrogate $\hat{\boldsymbol{\delta}}(P,t;\boldsymbol{\theta})\approx\boldsymbol{\delta}(P,t)$, with trainable parameters $\boldsymbol{\theta}$, over $P\in[P_{\min},P_{\max}]$ and $t\in[0,T]$, with horizon $T$, enabling rotor-angle prediction without numerical integration at inference. Following~\cite{misyris2020pinn}, the network takes the normalized input $\mathbf{x}^{(0)}=[\tilde P,\tilde t]^\top$, where $\tilde P=2\frac{P-P_{\min}}{P_{\max}-P_{\min}}-1$ and $\tilde t=2\frac{t}{T}-1$.

\section{Physics-Informed Networks}\label{sec:methods}
The standard PINN uses an MLP with $L$ hidden layers, each of width $N$, defined as
\begin{equation}
\begin{aligned}
\mathbf{x}^{(\ell)}&=\tanh\!\big(\mathbf{W}^{(\ell)}\mathbf{x}^{(\ell-1)}+\mathbf{b}^{(\ell)}\big),\quad \ell=1,\dots,L,\\
\hat{\boldsymbol\delta}&=\mathbf{W}^{(L+1)}\mathbf{x}^{(L)}+\mathbf{b}^{(L+1)},
\end{aligned}
\label{eq:mlp}
\end{equation}
where $\boldsymbol{\theta}=\{\mathbf{W}^{(\ell)},\mathbf{b}^{(\ell)}\}_{\ell=1}^{L+1}$ denotes the set of trainable parameters and $\tanh(\cdot)$ is the hidden-layer activation function (see Fig.~\ref{fig1}(a)). The c-PIKAN replaces linear transformations with learnable univariate functions associated with network edges, whose outputs are summed at each neuron~\cite{toscano2025pikan}. For a network with \(N_\ell\) nodes in layer \(\ell\), each layer is defined by
\begin{equation}
x_j^{(\ell)}=\sum_{i=1}^{N_{\ell-1}}\phi_{ij}^{(\ell)}\big(x_i^{(\ell-1)}\big),\quad
\phi_{ij}^{(\ell)}(x)=\sum_{k=0}^{K}c_{ijk}^{(\ell)}\,T_k(\tanh x),
\label{eq:kan}
\end{equation}
where $T_k$ is the first-kind Chebyshev polynomial, defined as
\begin{equation}
T_0(z)=1,\,\,T_1(z)=z,\,\, T_{k+1}(z)=2z\,T_k(z)-T_{k-1}(z),
\label{eq:cheb}
\end{equation}
where the polynomial degree is set to $K=5$ (see Fig.~\ref{fig1}(b)). The $\tanh(\cdot)$ transformation maps edge inputs to $[-1,1]$, the Chebyshev domain, while the trainable coefficients $c_{ijk}^{(\ell)}$ define the edge functions and constitute $\boldsymbol\theta$. The basis exhibits oscillatory behaviour, as implied by $T_k(\cos\vartheta)=\cos(k\vartheta)$, and its recursion in~\eqref{eq:cheb} requires only additions and multiplications, unlike the piecewise B--spline recursion of the bs--PIKAN~\cite{shuai2025pikan}.

The time derivatives of the network output are computed using automatic differentiation, with $\partial_t\hat\delta=\frac{2}{T}\partial_{\tilde t}\hat\delta$ and
$\partial_t^2\hat\delta=\frac{4}{T^2}\partial_{\tilde t}^2\hat\delta$. The physics residual for machine $i$ is defined as
\begin{equation}
R_i(P,t)=M_i\,\partial_t^2\hat\delta_i+D_i\,\partial_t\hat\delta_i-P_{m,i}(P)+P_{e,i}(\hat{\boldsymbol\delta}),
\label{eq:res}
\end{equation}
and all networks are trained by minimizing the loss adopted from~\cite{misyris2020pinn} (see Fig.~\ref{fig1}(c)):
\begin{equation}
\mathcal{L}=\underbrace{\frac{1}{N_u}\sum_{q=1}^{N_u}\|\hat{\boldsymbol\delta}(P_q,t_q)-\boldsymbol\delta_q\|^2}_{\mathrm{MSE}_u}
+\underbrace{\frac{1}{nN_f}\sum_{q=1}^{N_f}\sum_{i=1}^{n}R_i^2(P_q,t_q)}_{\mathrm{MSE}_f},
\label{eq:loss}
\end{equation}
where $N_u$ denotes the number of training data $(P_q,t_q,\boldsymbol{\delta}_q)$ sampled at $t=0$ and along the boundary trajectories at $P_{\min}$ and $P_{\max}$, while $N_f$ is the number of collocation points. For SMIB, $n=1$, and \eqref{eq:loss} reduces to the loss in~\cite{misyris2020pinn}. The loss enforces the initial angle through training data but does not explicitly constrain the initial angular speed $\boldsymbol{\omega}_0$, which is only indirectly enforced by the physics residual at interior operating points. An explicit penalty would require an additional weighting parameter, potentially exacerbating loss imbalance in multi-term objectives~\cite{xu2026review}.

\section{Proposed mac-PIKAN}\label{sec:mac}
Small-signal analysis provides a linear baseline capturing the dominant oscillatory modes, from which c-PIKAN learns the nonlinear correction. Let $\boldsymbol{\delta}_s(P)$ denote the stable post-disturbance equilibrium, assumed to exist (e.g., relative to an infinite bus or a reference machine), satisfying $P_{e,i}(\boldsymbol{\delta}_s)=P_{m,i}(P)$. Linearization of~\eqref{eq:mm} about this equilibrium yields $\mathbf{M}\Delta\ddot{\boldsymbol\delta}+\mathbf{D}\Delta\dot{\boldsymbol\delta}+\mathbf{K}\Delta\boldsymbol\delta=\mathbf 0$, where $\mathbf{M}=\mathrm{diag}(M_i)$, $\mathbf{D}=\mathrm{diag}(D_i)$, $K_{ij}=\left.\partial P_{e,i}/\partial\delta_j\right|_{\boldsymbol{\delta}_s}$, and $\Delta\boldsymbol{\delta}=\boldsymbol{\delta}-\boldsymbol{\delta}_s$. Introducing the state vector $\mathbf{z}=[\Delta\boldsymbol{\delta};\Delta\dot{\boldsymbol{\delta}}]$ gives $\dot{\mathbf{z}}=\mathbf{A}\mathbf{z}$, where
\begin{equation*}
\mathbf{A}=\begin{bmatrix}
    \mathbf{0}&\mathbf{I}\\-\mathbf{M}^{-1}\mathbf{K}&-\mathbf{M}^{-1}\mathbf{D}
\end{bmatrix}.
\end{equation*}
Assuming $\mathbf{A}$ is diagonalizable, $\mathbf{A}=\mathbf{V}\boldsymbol{\Lambda}\mathbf{V}^{-1}$, with eigenvalues $\lambda_r$, which for oscillatory modes form pairs $\sigma_r\pm j\omega_r$, the linearized response from the initial state $\mathbf{z}_0=[\boldsymbol{\delta}_0-\boldsymbol{\delta}_s;\boldsymbol{\omega}_0]$ is
\begin{equation}
\mathbf{z}(t)=\mathbf{V}\,\mathrm{diag}\big(e^{\lambda_rt}\big)\mathbf{V}^{-1}\mathbf{z}_0.
\label{eq:zlin}
\end{equation}
It comprises exponentially damped modal oscillations of the form $e^{\sigma_rt}\cos(\omega_rt)$ and $e^{\sigma_rt}\sin(\omega_rt)$. Let $\boldsymbol{\delta}_{\mathrm{lin}}=\boldsymbol{\delta}_s+\Delta\boldsymbol{\delta}(t)$ denote its rotor-angle component. For the SMIB system~\eqref{eq:smib}, the equilibrium angle, synchronizing coefficient, decay rate, and damped frequency are $\delta_s=\arcsin(P/B)$, $k_s=B\cos\delta_s$, $\sigma=-d/(2m)$, and $\omega_d=\sqrt{k_s/m-\sigma^2}$, respectively. Thus,
\begin{equation}
\delta_{\mathrm{lin}}=\delta_s+e^{\sigma t}\Big[\Delta_0\cos\omega_dt+\tfrac{\omega_0-\sigma\Delta_0}{\omega_d}\sin\omega_dt\Big],\quad
\Delta_0=\delta_0-\delta_s,
\label{eq:smiblin}
\end{equation}
where $P<B$ gives a stable equilibrium and $k_s/m>\sigma^2$ an underdamped response. The proposed mac-PIKAN is
\begin{equation}
\hat{\boldsymbol\delta}(P,t)=\boldsymbol\delta_{\mathrm{lin}}(P,t)+\Big(\frac{t}{T}\Big)^{2}\mathcal{N}_{\boldsymbol\theta}(P,t),
\label{eq:ps}
\end{equation}
where \(\mathcal{N}_{\boldsymbol{\theta}}\) denotes the c-PIKAN defined by~\eqref{eq:kan} (see Fig.~\ref{fig:smib}(b)). The linear response $\boldsymbol{\delta}_{\mathrm{lin}}$ is precomputed and held fixed during training. Only the coefficients $c_{ijk}^{(\ell)}$ are optimized using the loss in~\eqref{eq:loss}. Computing $\boldsymbol{\delta}_{\mathrm{lin}}$ requires only the system parameters and initial conditions already specified. Equation~\eqref{eq:ps} extends the trial solution of Lagaris \emph{et al.}~\cite{lagaris1998}, $\psi=A+Bt+t^2\mathcal{N}$, by replacing the polynomial component with the small--signal response. Specifically, $A=\boldsymbol{\delta}_{\mathrm{lin}}$ satisfies the initial conditions, while the correction $F_{\boldsymbol{\theta}}=h\,\mathcal{N}_{\boldsymbol{\theta}}$, with $h(t)=(t/T)^2$, and its first time derivative vanish at $t=0$. Thus, $\hat{\boldsymbol{\delta}}=A+F_{\boldsymbol{\theta}}$, $A:=\boldsymbol{\delta}_{\mathrm{lin}}$, and the exact nonlinear correction is $\mathbf{e}:=\boldsymbol{\delta}-A$.

\begin{theorem}[Trial solution]\label{thm}
\emph{(i)} For any $\boldsymbol\theta$ and $P$, the trial solution satisfies
$\hat{\boldsymbol\delta}(P,0)=\boldsymbol\delta_0$ and
$\partial_t\hat{\boldsymbol\delta}(P,0)=\boldsymbol\omega_0$. \emph{(ii)} Let $\Omega:=[P_{\min},P_{\max}]\times[0,T]$. If $\boldsymbol\delta\in C^2(\Omega)$, then
$\boldsymbol\delta(P,t)=A(P,t)+h(t)\mathbf r(P,t)$, where
$h(t)=(t/T)^2$ and $\mathbf r(P,t)=T^2\int_0^1(1-s)\,\ddot{\mathbf e}(P,st)\,ds$ is continuous on $\Omega$. By universal approximation~\cite{liu2024kan}, for any $\eta>0$ there exists a sufficiently large c-PIKAN with parameters $\boldsymbol\theta$ such that
\begin{equation}
\sup_{\Omega}\big\|\mathcal N_{\boldsymbol\theta}(P,t)-\mathbf r(P,t)\big\|<\eta,
\label{eq:thm_ua}
\end{equation}
and hence $\sup_{\Omega}\|\hat{\boldsymbol\delta}-\boldsymbol\delta\|<\eta$, since $0\le h\le 1$.
\end{theorem}

\begin{IEEEproof}
\emph{(i)} By \eqref{eq:zlin}, $\mathbf z(0)=\mathbf z_0$, so $A(P,0)=\boldsymbol\delta_0$ and $\partial_tA(P,0)=\boldsymbol\omega_0$; since $h(0)=\dot h(0)=0$, $F_{\boldsymbol\theta}$ and $\partial_tF_{\boldsymbol\theta}$ vanish at $t=0$. \emph{(ii)} By (i), $\mathbf e(P,0)=\dot{\mathbf e}(P,0)=\mathbf 0$, so Taylor's theorem with integral remainder gives
\begin{equation*}
\begin{aligned}
\mathbf e(P,t)
&=\mathbf e(P,0)+t\,\dot{\mathbf e}(P,0)
+t^2\!\int_0^1(1-s)\,\ddot{\mathbf e}(P,st)\,ds\\
&=t^2\!\int_0^1(1-s)\,\ddot{\mathbf e}(P,st)\,ds=h(t)\,\mathbf r(P,t),
\end{aligned}
\end{equation*}
with $\mathbf r$ continuous because $\ddot{\mathbf e}$ is continuous on the compact set $\Omega$. The bound \eqref{eq:thm_ua} follows from the density of KANs in $C(\Omega)$~\cite{liu2024kan}, and $\hat{\boldsymbol\delta}-\boldsymbol\delta=h\,(\mathcal N_{\boldsymbol\theta}-\mathbf r)$ with $0\le h\le1$ gives $\sup_{\Omega}\|\hat{\boldsymbol\delta}-\boldsymbol\delta\|\le\sup_{\Omega}\|\mathcal N_{\boldsymbol\theta}-\mathbf r\|<\eta$.
\end{IEEEproof}

\begin{proposition}\label{prop}
Suppose $P_e\in C^2$ on an open set containing $\{\boldsymbol\delta_s+\mathbf x:\|\mathbf x\|\le\gamma\}$ for some $\gamma>0$, and let $\epsilon:=\|\mathbf z_0\|$ be the initial deviation. Then there exist constants
$\bar\epsilon,C_e,C_r>0$ such that, for every $\epsilon\le\bar\epsilon$, the exact nonlinear correction in Theorem~\ref{thm} satisfies $\max_{t\in[0,T]}\|\mathbf e(P,t)\|\le C_e\,\epsilon^2$ and $\max_{t\in[0,T]}\|\mathbf r(P,t)\|\le C_r\,\epsilon^2$.
\end{proposition}

\begin{IEEEproof}
Let $\boldsymbol\rho(\mathbf x):=P_e(\boldsymbol\delta_s+\mathbf x)-P_e(\boldsymbol\delta_s)-\mathbf K\mathbf x$; since $P_e\in C^2$, Taylor's theorem gives $\|\boldsymbol\rho(\mathbf x)\|\le L_\rho\|\mathbf x\|^2$ for $\|\mathbf x\|\le\gamma$ and some $L_\rho>0$. Subtracting the linearized from the nonlinear swing equations, the error state $\mathbf y:=[\mathbf e;\dot{\mathbf e}]$ satisfies
\begin{equation*}
\dot{\mathbf y}=\mathbf A\mathbf y-\big[\mathbf 0;\,\mathbf M^{-1}\boldsymbol\rho(\Delta\boldsymbol\delta+\mathbf e)\big],\qquad \mathbf y(0)=\mathbf 0,
\end{equation*}
since $A$ and $\boldsymbol\delta$ share the initial conditions. Variation of constants gives $\|\mathbf y(t)\|\le c_T\max_{\tau\le t}\|\boldsymbol\rho(\Delta\boldsymbol\delta+\mathbf e)\|$, with $c_T$ depending only on $\mathbf A$, $\mathbf M$, and $T$. Let $\mathcal E(t):=\max_{\tau\le t}\|\mathbf y(\tau)\|$ and $C_{\mathrm{lin}}:=\max_{t\in[0,T]}\|e^{\mathbf At}\|$, so that $\|\Delta\boldsymbol\delta\|\le C_{\mathrm{lin}}\epsilon$ by \eqref{eq:zlin}, and choose $\bar\epsilon$ with $9c_TL_\rho C_{\mathrm{lin}}\bar\epsilon\le1$ and $3C_{\mathrm{lin}}\bar\epsilon\le\gamma$. For $\epsilon\le\bar\epsilon$, if $\mathcal E(t)\le2C_{\mathrm{lin}}\epsilon$, then $\|\Delta\boldsymbol\delta+\mathbf e\|\le3C_{\mathrm{lin}}\epsilon\le\gamma$ and thus
\begin{equation*}
\mathcal E(t)\le c_TL_\rho(3C_{\mathrm{lin}}\epsilon)^2=9c_TL_\rho C_{\mathrm{lin}}^2\epsilon^2\le C_{\mathrm{lin}}\epsilon;
\end{equation*}
since $\mathcal E(0)=0$ and $\mathcal E$ is continuous, this holds on $[0,T]$, giving $\|\mathbf e\|,\|\dot{\mathbf e}\|\le C_e\epsilon^2$ with $C_e:=9c_TL_\rho C_{\mathrm{lin}}^2$. Finally, $\mathbf M\ddot{\mathbf e}=-\mathbf D\dot{\mathbf e}-\mathbf K\mathbf e-\boldsymbol\rho(\Delta\boldsymbol\delta+\mathbf e)$ yields $\|\ddot{\mathbf e}\|\le C'\epsilon^2$ for an $\epsilon$-independent $C'$, and by Theorem~\ref{thm}(ii), $\|\mathbf r\|\le T^2\int_0^1(1-s)\|\ddot{\mathbf e}\|\,ds\le C_r\epsilon^2$ with $C_r:=\tfrac{T^2}{2}C'$.
\end{IEEEproof}

\begin{remark}\label{rem:interpretation}
\emph{(i)} Proposition~\ref{prop} establishes an $O(\epsilon^2)$ correction for small disturbances; Section~\ref{sec:results} confirms this scaling numerically and shows that the correction is learned effectively beyond this regime. \emph{(ii)} Replacing the c-PIKAN in~\eqref{eq:ps} with an MLP gives a modal-augmented PINN (ma-PINN), which also satisfies both initial conditions exactly and serves as an ablation separating the modal augmentation from the network.
\end{remark}

\section{Benchmark and Evaluation Protocol}\label{sec:setup}
Table~\ref{tab:setup} summarizes the benchmark settings from~\cite{misyris2020pinn} and the evaluated model configurations. We reproduce the benchmark in PyTorch using the original random seed to generate identical training data and collocation points for all models\footnote{All runs: AMD Ryzen 9 5900HS (8 cores, 32\,GB RAM), Windows 11, Python 3.12, PyTorch 2.14.1 on CPU, one thread each, six in parallel.}. Models use $N_f=8000$ collocation points, and errors are evaluated against the reference solution at 100 values of $P$ and 201 time points with $\Delta t=0.1$~s. All models are trained with L--BFGS--B (max 50{,}000 iterations, history size 50, convergence tolerance $10^{-15}$). Trainable parameters comprise one weight per connection (plus biases) for the PINN, $K+1=6$ Chebyshev coefficients for the c--PIKAN, and 13 B-spline coefficients, one base weight, and one scale parameter for the bs--PIKAN. Five random initializations are used because the TensorFlow initialization in~\cite{misyris2020pinn} cannot be reproduced in PyTorch. The relative $L_2$ error is $\varepsilon=\|\hat{\mathbf y}-\mathbf y\|_2/\|\mathbf y\|_2\times100\%$, where $\hat{\mathbf y}$ and $\mathbf y$ are predicted and reference outputs. Errors are computed on the test grid for the rotor angle and the frequency $\omega=\dot{\delta}$, obtained by numerical differentiation as in~\cite{misyris2020pinn}.

\begin{table}[!t]
\centering
\caption{Benchmark setting of~\cite{misyris2020pinn} (top) and models (bottom)}
\label{tab:setup}
\scriptsize\setlength{\tabcolsep}{2.6pt}
\renewcommand{\arraystretch}{1.2}
\begin{tabular*}{\columnwidth}{@{\extracolsep{\fill}}ccccccccc}
\toprule
\multicolumn{4}{c}{system \eqref{eq:smib}} & \multicolumn{2}{c}{domain} & \multicolumn{3}{c}{data} \\
\cmidrule(lr){1-4}\cmidrule(lr){5-6}\cmidrule(lr){7-9}
$m$ & $d$ & $B$ & $\delta_0=\omega_0$ & $P$ [p.u.] & $t$ [s] & test grid $P\times t$ & $N_u$ & $N_f$ \\
\midrule
0.4 & 0.15 & 0.2 & 0.1 & $[0.08,0.18]$ & $[0,20]$ & $100\times201$ & 40 & 8000 \\
\bottomrule
\end{tabular*}

\vspace{4pt}
\setlength{\tabcolsep}{1pt}
\begin{tabular*}{\columnwidth}{@{\extracolsep{\fill}}lcccccc}
\toprule
 & hidden layers & edge & connec- & per con- & para- & modal \\
 & $\times$ nodes & function & tions & nection & meters & term \eqref{eq:smiblin} \\
\midrule
PINN~\cite{misyris2020pinn} & $5\times10$ & $\tanh$ (fixed) & 460 & 1 & 481$^{*}$ & -- \\
bs-PIKAN~\cite{shuai2025pikan} & $1\times5$ & B-spline & 15 & 15 & 225 & -- \\
c-PIKAN & $2\times8$ & Chebyshev & 88 & 6 & 528 & -- \\
ma-PINN & $5\times10$ & $\tanh$ (fixed) & 460 & 1 & 481$^{*}$ & \checkmark \\
mac-PIKAN & $2\times8$ & Chebyshev & 88 & 6 & 528 & \checkmark \\
\bottomrule
\end{tabular*}
\par\vspace{2pt}\noindent\parbox{\columnwidth}{\scriptsize $^{*}$460 weights and 21 biases.}
\end{table}

\section{Results}\label{sec:results}
\subsection{Accuracy}
\begin{figure}[!t]
\centering
\subfloat[$P=0.180$\,p.u.\label{fig:traja}]{\includegraphics[width=0.49\columnwidth]{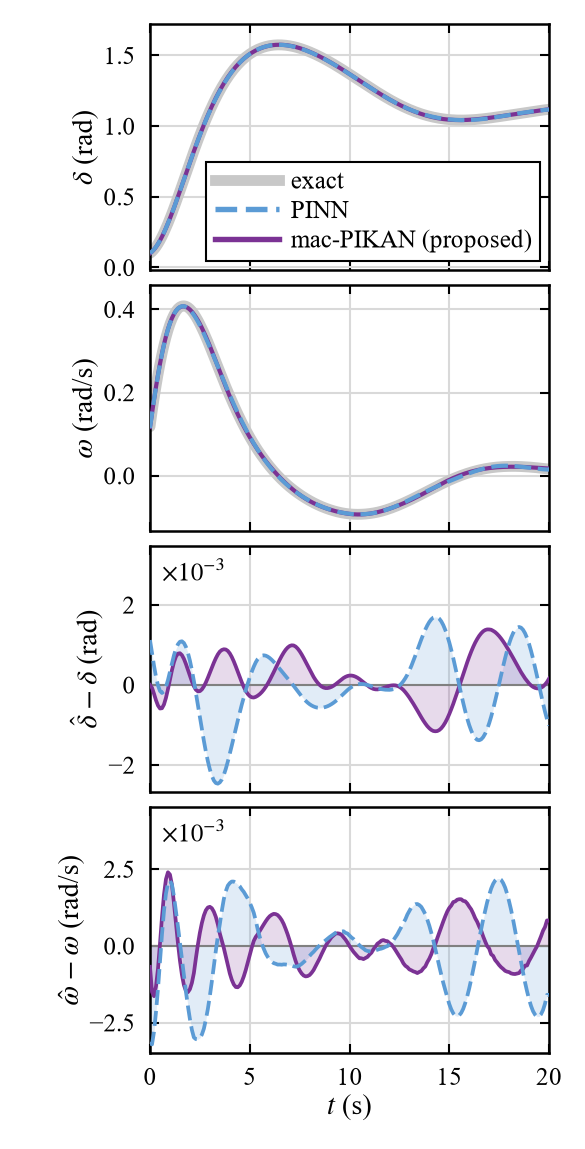}}\hfill
\subfloat[$P=0.163$\,p.u.\label{fig:trajb}]{\includegraphics[width=0.49\columnwidth]{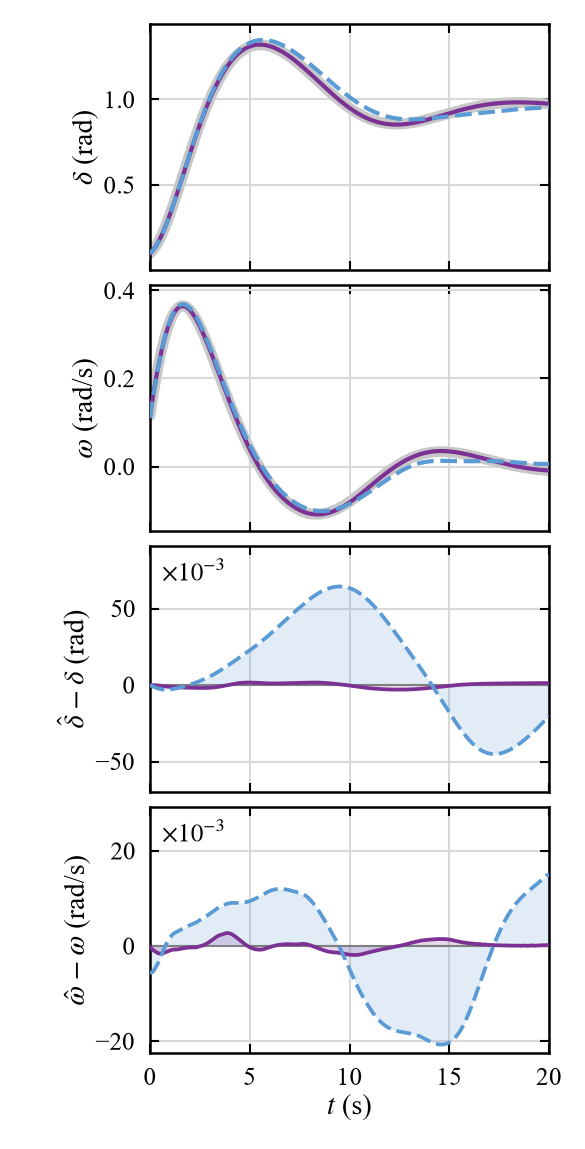}}
\caption{SMIB trajectories for seed 0 with the lowest (a) and highest (b) PINN error: rotor angle and frequency (top) and their errors over time (bottom).}
\label{fig:traj}
\end{figure}

Table~\ref{tab:bench} lists the rotor-angle and frequency errors. The small-signal model~\eqref{eq:smiblin} alone yields errors of 6.39\% and 15.26\%, while mac-PIKAN achieves the lowest mean errors in both (0.29\% and 0.99\%)---5.5 and 5.3 times lower than the PINN (1.59\% and 5.26\%). mac-PIKAN outperforms the PINN on every seed, with its worst-seed error (0.39\%) remaining below the PINN mean. All models fit the boundary trajectories to below $10^{-3}$. The bs-PIKAN model ranks second in angle accuracy (0.86\%) but requires 6.7 times longer to train. ma-PINN (Remark~\ref{rem:interpretation}(ii)) also improves on the PINN, particularly in frequency (2.77 versus 5.26\%), so modal augmentation benefits the MLP as well. For seed~0, the PINN errors (0.08--3.70\%) are comparable to those in~\cite{misyris2020pinn} (0.03--2.37\%): on its best trajectory (Fig.~\ref{fig:traj}(a)) all models nearly coincide with the exact solution, whereas on its worst (Fig.~\ref{fig:traj}(b)) its angle error reaches 0.064~rad, while mac-PIKAN satisfies the initial condition exactly and remains within 0.003~rad.

\begin{table}[!t]
\centering
\caption{Errors on the test grid (mean\,$\pm$\,std, five seeds); bold: best}
\label{tab:bench}
\setlength{\tabcolsep}{2.6pt}
\renewcommand{\arraystretch}{1.2}
\resizebox{\columnwidth}{!}{\begin{tabular}{lccccccc}
\toprule
 & Para- & \multicolumn{4}{c}{relative $L_2$ error [$\times10^{-2}$]} & max.\ $\omega(0)$ & time \\
\cmidrule(lr){3-6}
 & meters & angle $\delta$ & frequency $\omega$ & best traj. & worst traj. & error [rad/s] & [s] \\
\midrule
Small-signal model \eqref{eq:smiblin} & 0 & 6.39 & 15.26 & 0.59 & 12.36 & 0.009 & -- \\
\midrule
PINN~\cite{misyris2020pinn} & 481 & 1.59\,$\pm$\,0.71 & 5.26\,$\pm$\,2.30 & 0.07 & 2.26 & 0.024 & 82 \\
bs-PIKAN~\cite{shuai2025pikan} & 225 & 0.86\,$\pm$\,0.41 & 3.74\,$\pm$\,1.79 & \best{0.02} & 1.53 & 0.024 & 4368 \\
c-PIKAN & 528 & 3.14\,$\pm$\,1.30 & 13.58\,$\pm$\,5.52 & \best{0.02} & 5.28 & 0.098 & 634 \\
ma-PINN & 481 & 0.97\,$\pm$\,0.34 & 2.77\,$\pm$\,0.69 & 0.09 & 1.54 & \best{0.001} & 89 \\
\rowcolor{black!6}mac-PIKAN & 528 & \best{0.29\,$\pm$\,0.06} & \best{0.99\,$\pm$\,0.24} & 0.04 & \best{0.54} & \best{0.001} & 656 \\
\bottomrule
\end{tabular}}
\end{table}

\subsection{Initial Conditions}
\begin{figure}[!t]
\centering
\subfloat[Training loss\label{fig:lossma}]{\includegraphics[width=0.49\columnwidth]{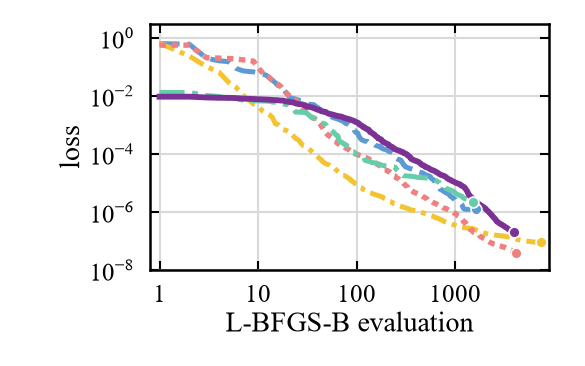}}\hfill
\subfloat[Angle error over $P$\label{fig:lossmb}]{\includegraphics[width=0.49\columnwidth]{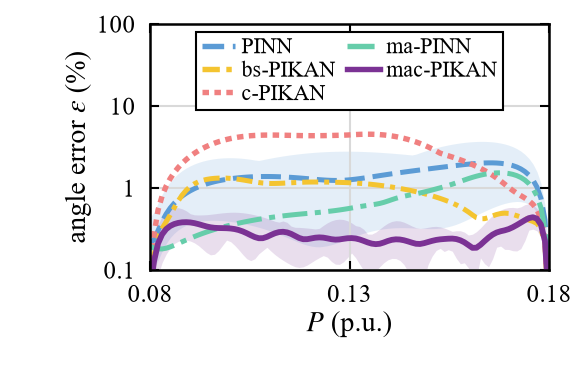}}
\caption{Five models over five seeds; (a) median, dots: convergence; (b) mean, shaded: seed range of PINN and mac-PIKAN.}
\label{fig:lossm}
\end{figure}

As seen in Fig.~\ref{fig:lossm}(a), the plain c-PIKAN reaches the lowest training loss, yet Table~\ref{tab:bench} shows that it has the largest prediction error. Only 8 of the 40 training data lie at $t=0$; they constrain the initial angle but not its time derivative, so the initial speed of c-PIKAN deviates at interior $P$ and produces a different trajectory that still satisfies the residual. In contrast, the proposed method enforces both $\delta(0)$ and $\omega(0)$ exactly for every $P$ (see Theorem~\ref{thm}) and starts training at a lower loss ($10^{-2}$ versus $0.6$). The network learns only the correction in Proposition~\ref{prop}, numerically confirming its $\epsilon^2$ scaling (see Remark~\ref{rem:interpretation}(i)) and accurate learning at $\epsilon=1.02$. The Chebyshev basis better represents the correction than the MLP, with Fig.~\ref{fig:lossm}(b) showing the largest PINN improvement midway between boundary trajectories, far from training data. The bs--PIKAN model also avoids the initial--speed error, likely owing to its smoothly interpolating local B--spline basis.

\subsection{Robustness and Training Cost}
\begin{figure}[!t]
\centering
\subfloat[Training data $N_u$\label{fig:robusta}]{\includegraphics[width=0.49\columnwidth]{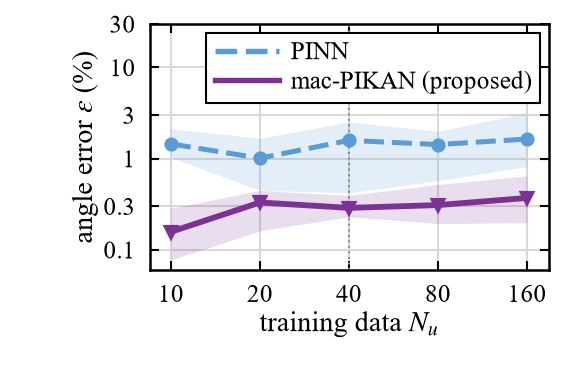}}\hfill
\subfloat[Damping $d$\label{fig:robustb}]{\includegraphics[width=0.49\columnwidth]{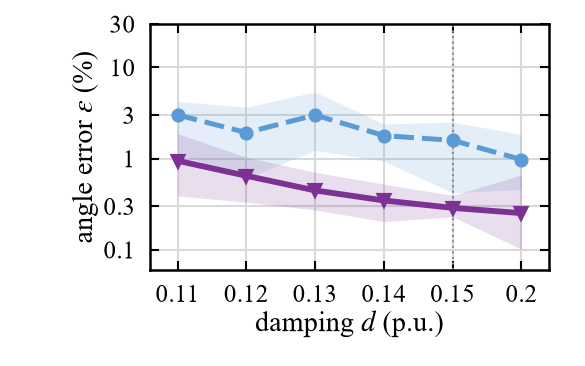}}
\caption{Angle error over all trajectories when one setting of~\cite{misyris2020pinn} is changed; lines: mean, bands: range of five seeds, dotted: original setting.}
\label{fig:robust}
\end{figure}

\begin{figure}[!t]
\centering
\subfloat[Worst for PINN\label{fig:d011a}]{\includegraphics[width=0.49\columnwidth]{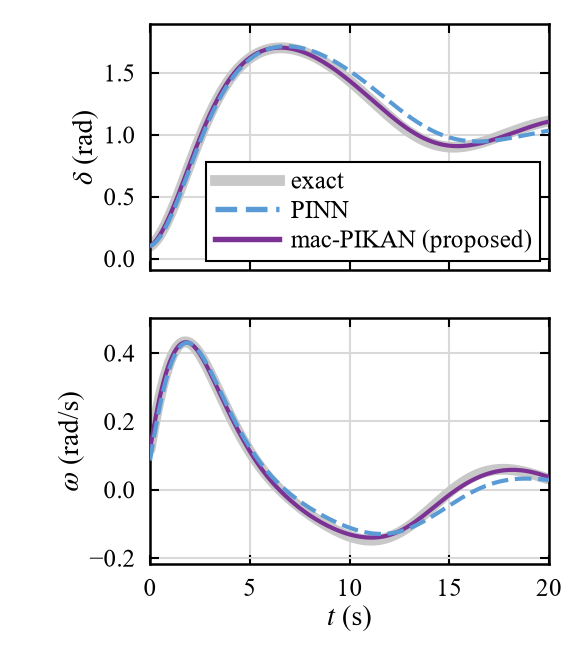}}\hfill
\subfloat[Worst for mac-PIKAN\label{fig:d011b}]{\includegraphics[width=0.49\columnwidth]{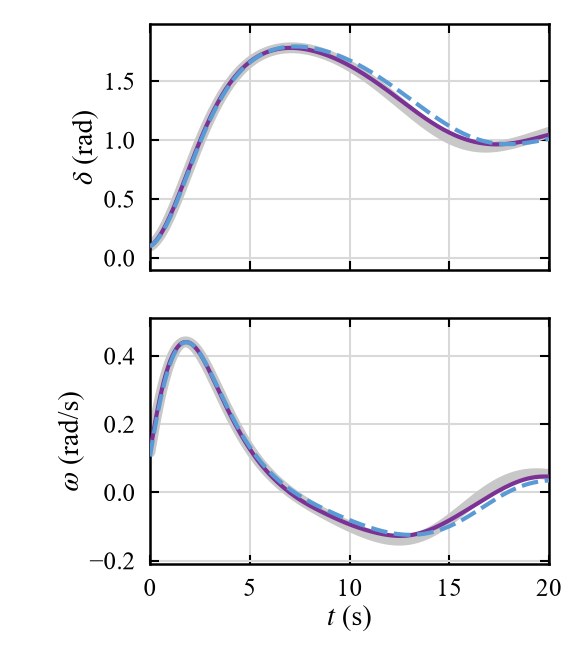}}
\caption{Trajectories at $d=0.11$, all other settings as in~\cite{misyris2020pinn}, for the run with the median PINN error of
five seeds ($P=0.175$ and 0.178\,p.u.).}
\label{fig:d011}
\end{figure}

Figure~\ref{fig:robust} varies one parameter of the model in~\cite{misyris2020pinn} at a time\footnote{For parameter variations from~\cite{misyris2020pinn}, reference trajectories and training data are recomputed by numerically integrating~\eqref{eq:smib} on the same $100\times201$ grid.}. As Fig.~\ref{fig:robust}(a) shows, more training data barely change the PINN error, since boundary data say little about the interior, whereas mac-PIKAN is 3 to 9 times more accurate, most with few data. In Fig.~\ref{fig:robust}(b), lowering the damping from 0.2 to 0.11 degrades both models, yet mac-PIKAN stays 3 to 7 times more accurate, and at $d=0.12$ it is better on four of five seeds. Fig.~\ref{fig:d011} shows the run with the median PINN error at $d=0.11$: on its hardest trajectory (Fig.~\ref{fig:d011}(a)), the PINN drifts after the first swing, with 5.30\% angle error against 0.65\% for mac-PIKAN, and even on the hardest trajectory of mac-PIKAN, next to $P_{\max}$ (Fig.~\ref{fig:d011}(b)), where by Proposition~\ref{prop} the correction is largest, it remains 2.6 times more accurate. Table~\ref{tab:bench} gives the training times: one c-PIKAN evaluation is 4.4 times cheaper than a bs-PIKAN one, and mac-PIKAN trains 6.7 times faster than bs-PIKAN, though about 8 times slower than the PINN. More computation does not close this gap: with $N_f=80{,}000$, the PINN needs 1176~s and still reaches 1.00\%, whereas a smaller mac-PIKAN ($2\times7$, 420 parameters, fewer than the PINN) reaches 0.32\% in about half its training time.

\section{Conclusion}
This paper investigated Chebyshev-based PIKANs for power system dynamic simulation and proposed mac--PIKAN, which incorporates the closed--form response of the linearized swing equations. While c--PIKANs are cheaper to train than bs-PIKANs, they leave the initial speed unconstrained; mac--PIKAN overcomes this limitation by enforcing both initial conditions exactly and learning only the nonlinear correction. On the single-machine infinite-bus benchmark, mac--PIKAN achieves the highest accuracy among all tested models, particularly between boundary trajectories, and maintains this advantage across training data sizes and at lower damping, albeit at a higher training cost than PINNs. Future work will extend the framework to larger power systems, integrate parameter identification, reduce training cost, and develop time-marching strategies for long-horizon prediction.

\bibliographystyle{IEEEtran}
\bibliography{refs}
\end{document}